# Unique Sense: Smart Computing Prototype

Vijaykumar S, Saravanakumar S.G, Dr. M. Balamurugan

*Student, School of Computer Science, Engineering and Applications, Bharathidasan University, Trichy – 23, TamilNadu, India*
*indianid@gmail.com*
*Member Technical Staff, 6th SENSE, An Advanced Research and Scientific Experiment Foundation, TamilNadu, India.*
*saravanakumarsg@gmail.com*
*Associate Professor, School of Computer Science, Engineering and Applications, Bharathidasan University, Trichy – 23, TamilNadu, India*
*mmbalmurugan@gmail.com,*

**Abstract**

Unique sense: Smart computing prototype is a part of "unique sense" computing architecture, which delivers alternate solution for today's computing architecture. This computing is one step towards future generation needs, which brings extended support to the ubiquitous environment. This smart computing prototype is the light weight compact architecture which is designed to satisfy all the needs of this society. The proposed solution is based on the hybrid combination of cutting edge technologies and techniques from the various layers. In addition it achieves low cost architecture and eco-friendly to meet all the levels of people's needs.

## 1. Introduction

We are living in the era where data is more important in deciding the future systems and its size grows every day. The scientists and other users who work on the complex problems perform daily operation with these large data set. These large data set creates huge load to the server. Even the super computer encounters difficulty when it involves large dataset: especially if the data are distributed across the network.

In this paper, we have taken our first step by creating new SC (Smart Computing) Prototype based on the ARM architecture which also solves the current issues like complex problems on the large data sets. The proposed system way also support the task oriented system and it will focus on the data loads which are the main issue of the HPC and other Cloud based DFS systems. This smart architecture works on the low powered ARM based chip (Raspberry pi) to reduce the power consumption, highly portable and it works along with the Hadoop Framework to support fault tolerance, data availability and easy computation in working with the large data set.

*1.1. HPC Systems*

High Performance Computing (HPC) is used for processing the complex data and it works in parallel on the distributed environments to produce effective and faster result. It is efficient in solving complex programs, Scientific, Weather forecasting systems, Medical systems, aerodynamic simulations, etc.

HPC systems are divided into two main categories vise., Shared memory system where all disk share the same memory space with multi-core or multiple processor and Cluster systems where all the individual systems are connected to a common network and works for the common problem [14] [16].

Task parallel HPC systems are primarily focuses on the scientific data set, complex problems and it gives secondary focus on the data loads [16].

*1.2. Cluster Cloud system*

Cluster System works on the cloud environments or shared networks, it serves to retrieve the static/ dynamic web pages, web services, and conventional distributed file system applications, where the data are taken as priority requirements. These systems are mainly designed to store the data effectively rather than executing on it [16].

**2. Problem Statement**

Based on the recent study the HPC systems facing number of limitations and the solutions are being applied in the industries or work environment based on the requirement. Here we address few limitations of HPC systems, To begin with the power consumption, generally power consumed in HPC system are huge when compared to cluster system it is due to the maximum utilization of hardware when it tends to process the complex operation. Infra Maintenance for the HPC System would be difficult because the heat generated by the multi-core processor is very high and it requires ultra-cooling fans and very good infra support to maintain the system temperature. HPC System are huge and it cannot be taken anywhere and installed to resolve the high complex problems. Memory needs to be increased in order to accommodate the increased data size and it results in the installation of more multi-core processor but it end up in performance degradation in the I/O transmission when the messages are getting transferred. HPC systems are designed in such a way where the user has more responsibilities to manage the HPC system [14] [15].

On the other part we try to address some of the issues found in the cloud systems, these systems serves for web pages, web services and other applications in the internet. Cluster Systems are not designed to work effectively on high complex problem. Cluster System work on the basis of scattered memory so the files are split in to the various systems and for complex problem the data need to be process in parallel on the multi-processor which in turn increases the rate of data exchange. The cost setup to maintain the common network is quite high and it cost a lot to exchange the data's for analyzing the problems to the required systems. [17]

**3. Challenges**

Identify the light weighted and low power consuming architecture that supports Hadoop installation and it should meet basic hardware requirements. The System should bring high availability, durability, fault tolerance, supports variety and high data loads. Need to identify whether the chosen architecture supports clustering and utilize maximum resources to contribute in improving the system performance. The system should utilize minimum amount of instructions set that can be handled by the single board computer. Identify the source of power and power consumption with the heat resistance.

## 4. Prototype

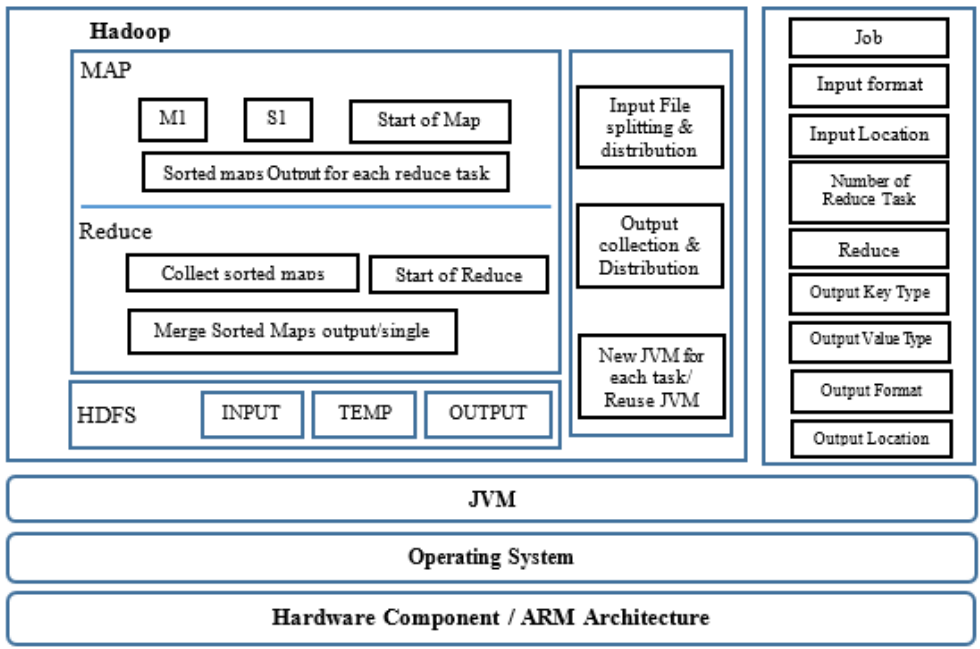

Fig 1. Smart Computing Architecture

- M1 – Map Algorithm for shorting the input
- S1 – Shuffle, Partition/short per map output

## 5. Hybrid Solution

*5.1 ARM*

ARM is also an instruction set architectures used by processors depend on RISC architecture. It represent three cortex profiling for Application, Real-time, Microcontroller known as Cortex A, Cortex R, and Cortex M. Mainly Raspberry pi comes with ARM1176JZ-F undefined series but most of the properties are as same as ARM 11 which is 32 bit ARM architecture, ARMv6 Architecture core. Especially those architectures are emit reduced heat when compare with previous models and lower heat risk and most compactable for real time process. Because most of the mobile phones are using this architecture. Especially Series 1176 having security extensions [9].

*5.2 Hadoop & Map reduce*

Hadoop is a platform that provides both distributed storage and computational capabilities. It brings support in two dimensions viz., HDFS for storage and map reduce for computational capabilities [7].

MapReduce is a Programming model and an associated implementation for processing and generating large data sets [4]. Users specify the computation in terms of a map and a reduce function, and the underlying runtime system automatically parallelizes the computation across large-scale clusters of machines, handles machine failures, and schedules inter-machine communication to make efficient use of the network and disks. Programmers find the system easy to use: more than ten thousand distinct mapreduce programs have been implemented internally at Google over the past four years, and an average of one hundred thousand mapreduce jobs are executed on Google's clusters every day, processing a total of more than twenty petabytes of data per day[6][4].

## 6. Implementation

### 6.1. Raspberry - Pi

Raspberry Pi Foundation is an educational charity situated in UK which have a motto in finding of advance education system and technology for society. As their contribution they developed Credit card size light weight processing computer called Raspberry Pi [11].

### 6.2. POWER SOURCE:

There are two possible way to provide power source for our system. In this architecture we have chosen micro USB instead of GIPO for achieving quick stability based on available resource having capability for providing power to I/O components. 2A - 5 V is the power factor meets our requirement.

### 6.3. OPERATING SYSTEM:

Code named wheezy is the one of the stable version from Debian, Linux distribution. With the future of multi-arch which support 32 bit runs on 64 bit operating system and its feature extends to support arm [8]. So here in this work we choose it as a one of the supporting system for Hadoop in ARM architecture. Therefore, we utilized Rasbian, Debian wheezy linux operating system Kernel Version 3.12, and released date 9 September 2014 from the Raspberry supporting site.

### 6.4. JVM:

ARM architecture supports Java environment and we need JVM for installing Hadoop framework. So we have installed java version "1.7.0_07" on the Raspberry Pi.

### 6.5. Hduser:

We have added sudo user having rights to install applications and that user will later added into Hadoop group to access the file system.

### 6.6. SSH Secure shell:

We have used SSH secure shell which are widely used protocol to connect remote system. After Creating SSH key share that key with user to establish communication between its nodes.

### 6.7. Permission:

The common type of permission where given to the users so that users can possibly process, read, write and execute (Traverse for directories)

### 6.8. Processes Invocation:

After installing the required component in Linux, most commonly we need to start the process manually

## 7. RESULT

hduser@raspberrypi /usr/local/hadoop $ hadoop jar hadoop-examples-1.1.2.jar pi 5 50
Number of Maps  = 5
Samples per Map = 50

Wrote input for Map #0
Wrote input for Map #1
Wrote input for Map #2
Wrote input for Map #3
Wrote input for Map #4
Starting Job
14/12/13 11:48:09 INFO mapred.FileInputFormat: Total input paths to process : 5
14/12/13 11:48:21 INFO mapred.JobClient: Running job: job_201412131131_0001
14/12/13 11:48:22 INFO mapred.JobClient: map 0% reduce 0%
14/12/13 11:52:33 INFO mapred.JobClient: map 20% reduce 0%
14/12/13 11:55:36 INFO mapred.JobClient: map 40% reduce 0%
14/12/13 11:55:49 INFO mapred.JobClient: map 40% reduce 13%
14/12/13 11:57:43 INFO mapred.JobClient: map 60% reduce 13%
14/12/13 11:57:55 INFO mapred.JobClient:  map 60% reduce 20%
14/12/13 11:58:04 INFO mapred.JobClient:  map 80% reduce 20%
14/12/13 11:58:18 INFO mapred.JobClient:  map 80% reduce 26%
14/12/13 11:59:15 INFO mapred.JobClient:  map 100% reduce 26%
14/12/13 11:59:27 INFO mapred.JobClient:  map 100% reduce 33%
14/12/13 11:59:46 INFO mapred.JobClient:  map 100% reduce 100%
14/12/13 12:00:40 INFO mapred.JobClient: Job complete: job_201412131131_0001
14/12/13 12:00:41 INFO mapred.JobClient: Counters: 30
14/12/13 12:00:42 INFO mapred.JobClient:   Job Counters
14/12/13 12:00:42 INFO mapred.JobClient: Launched reduce tasks=1
14/12/13 12:00:42 INFO mapred.JobClient: SLOTS_MILLIS_MAPS=976959
14/12/13 12:00:42 INFO mapred.JobClient: Total time spent by all reduces waiting after reserving slots (ms)=0
14/12/13 12:00:42 INFO mapred.JobClient: Total time spent by all maps waiting after reserving slots (ms)=0
14/12/13 12:00:42 INFO mapred.JobClient: Launched map tasks=6
14/12/13 12:00:42 INFO mapred.JobClient: Data-local map tasks=6
14/12/13 12:00:42 INFO mapred.JobClient: SLOTS_MILLIS_REDUCES=421500
14/12/13 12:00:42 INFO mapred.JobClient:   File Input Format Counters
14/12/13 12:00:42 INFO mapred.JobClient: Bytes Read=590
14/12/13 12:00:42 INFO mapred.JobClient: File Output Format Counters
14/12/13 12:00:42 INFO mapred.JobClient: Bytes Written=97
14/12/13 12:00:42 INFO mapred.JobClient: FileSystemCounters
14/12/13 12:00:42 INFO mapred.JobClient:
FILE_BYTES_READ=116
14/12/13 12:00:42 INFO mapred.JobClient: HDFS_BYTES_READ=1210
14/12/13 12:00:42 INFO mapred.JobClient: FILE_BYTES_WRITTEN=305403
14/12/13 12:00:42 INFO mapred.JobClient: HDFS_BYTES_WRITTEN=215
14/12/13 12:00:42 INFO mapred.JobClient: Map-Reduce Framework
14/12/13 12:00:42 INFO mapred.JobClient: Map output materialized bytes=140
14/12/13 12:00:42 INFO mapred.JobClient: Map input records=5
14/12/13 12:00:42 INFO mapred.JobClient: Reduce shuffle bytes=140
14/12/13 12:00:42 INFO mapred.JobClient: Spilled Records=20
14/12/13 12:00:42 INFO mapred.JobClient: Map output bytes=90
14/12/13 12:00:42 INFO mapred.JobClient: Total committed heap usage (bytes)=1021792256
14/12/13 12:00:42 INFO mapred.JobClient: CPU time spent (ms)=73380
14/12/13 12:00:42 INFO mapred.JobClient: Map input bytes=120
14/12/13 12:00:42 INFO mapred.JobClient: SPLIT_RAW_BYTES=620
14/12/13 12:00:42 INFO mapred.JobClient:    Combine input records=0
14/12/13 12:00:42 INFO mapred.JobClient:    Reduce input records=10

| | |
|---|---|
| 14/12/13 12:00:42 INFO mapred.JobClient: | Reduce input groups=10 |
| 14/12/13 12:00:42 INFO mapred.JobClient: | Combine output records=0 |
| 14/12/13 12:00:42 INFO mapred.JobClient: | Physical memory (bytes) snapshot=726032384 |
| 14/12/13 12:00:42 INFO mapred.JobClient: | Reduce output records=0 |
| 14/12/13 12:00:42 INFO mapred.JobClient: | Virtual memory (bytes) snapshot=2161262592 |
| 14/12/13 12:00:42 INFO mapred.JobClient: | Map output records=10 |

Job Finished in 758.142 seconds
Estimated value of Pi is 3.14800000000000000000

**Conclusion & Future works:**

We have described Smart computing prototype in this work and the proposed system is the part of our ongoing computing research project called "unique sense". Here the result shows that the successful execution of proposed prototype. This single node hybrid architecture is the initiation towards clustered provisioning and for the creation of models based on the dependencies. We are in the process of deploying the multi-node cluster to evaluate the performance and to fix the benchmarking from various perspective. Ultimately this architecture brings fault tolerance and reliable solution for the today's findings.

**Acknowledgements**

Authors Vijaykumar S. and Saravanakumar S. G. are Grateful to Research, Technical and Advisor members of 6[th] SENSE Research Foundation, Tamilnadu, South India. Url : www.6thsense.us